\documentclass{Interspeech2024}
\usepackage{bm}
\usepackage{tabularx}
\usepackage{multirow}
\newcommand\blfootnote[1]{%
  \begingroup
  \renewcommand\thefootnote{}\footnote{#1}%
  \addtocounter{footnote}{-1}%
  \endgroup
}

\interspeechcameraready

\begin{document}
\title{Interpretable Temporal Class Activation Representation for Audio Spoofing Detection}

\name{Menglu}{Li$^{1}$}
\name{Xiao-Ping}{Zhang$^{2,*}$}


\address{$^1$Toronto Metropolitan University, Canada\\$^2$Shenzhen Key Laboratory of Ubiquitous Data Enabling, Tsinghua Shenzhen International Graduate School, Tsinghua University, China}
\email{menglu.li@torontomu.ca, xpzhang@ieee.org}


\newcommand{\red}[1]{\textcolor{red}{#1}}

\maketitle

\begin{abstract}
    
Explaining the decisions made by audio spoofing detection models is crucial for fostering trust in detection outcomes. However, current research on the interpretability of detection models is limited to applying XAI tools to post-trained models. In this paper, we utilize the wav2vec 2.0 model and attentive utterance-level features to integrate interpretability directly into the model's architecture, thereby enhancing transparency of the decision-making process. Specifically, we propose a class activation representation to localize the discriminative frames contributing to detection. Furthermore, we demonstrate that multi-label training based on spoofing types, rather than binary labels as bonafide and spoofed, enables the model to learn distinct characteristics of different attacks, significantly improving detection performance. Our model achieves state-of-the-art results, with an EER of 0.51\% and a min t-DCF of 0.0165 on the ASVspoof2019-LA set.
\end{abstract}

\section{Introduction}
\blfootnote{$^*$X.-P. Zhang is the corresponding author.}Audio spoofing detection techniques have gained more attention recently due to the threat brought to automatic speaker verification (ASV) systems and their harmful impact of spoofed audio on societies. Spoofing countermeasures attempt to improve detection performance by developing both classifier structure and feature extraction methods. The detection algorithms usually operate upon the hand-crafted acoustic feature \cite{luo2021capsule, li2022role, li2023robust} until the emergence of advanced learnable front-ends, involving convolution-based networks \cite{jung2022aasist,tak2021end} or self-supervised learning (SSL)-based architectures \cite{wang2021investigating,tak2022automatic}, which outperform the traditional feature extraction. Regarding architectures of the classifier, Deep learning (DL)-based countermeasures have demonstrated their advance compared to traditional machine learning models, such as GMM \cite{li2022comparative}. Notably, light CNN \cite{ma2021improved,tomilov2021stc}, ResNet \cite{hua2021towards,li2021replay}, and DARTS \cite{wang2022fully,ge21_asvspoof} algorithms have made great achievements in increasing the detection accuracy on both known and unknown spoofing attacks generated by Text-to-Speech (TTS) and Voice Conversion (VC) techniques.

While current state-of-the-art models achieve high accuracy on publicly available datasets, they fail to provide explanations for their detection outcomes or the decision-making process due to the black-box nature of DL-based techniques. Designing a detection algorithm with higher interpretability, especially through visualization, is crucial. It not only clarifies how and why a detection model makes decisions, fostering trust in the algorithm and its outputs, but also allows for understanding the architecture of the algorithm and potential improvements by adjusting specific segments. To address this problem, we propose an interpretable audio spoofing detection model with a guaranteed high detection rate. We utilize a pre-trained SSL model to extract frame-level features, which are then combined with utterance-level information to form a high-level embedding of the audio input. Our proposed detection pipeline incorporates an attention mechanism along feature channels to generate a temporal class activation representation. This representation effectively localizes the most discriminative frames contributing to different labels in the detection process, while also making them visually accessible.

The new contributions in this work are (1) We propose a novel audio spoofing detection model that leverages an effective feature set comprising SSL-based frame-level features and attentive utterance-level features. (2) The proposed model provides a class activation map as a visualizable interpretation of detection results, revealing the underlying temporal dynamics. (3) We demonstrate the effectiveness of employing multi-label classification training, rather than binary labels, to learn distinct characteristics of TTS and VC-based artifacts.

\section{Related Work}
A group of existing works have utilized explainable artificial intelligence (XAI) tools to uncover the behaviour of deep neural network algorithms in detecting spoofed audio \cite{li2024audio}. Ge et al. \cite{ge22_odyssey} utilize SHapley Additive exPlanations (SHAP) \cite{lundberg2017unified} to identify the characteristics of artifacts relied on various spoofing attacks. Lim et al. \cite{lim2022detecting} apply both Deep Taylor \cite{montavon2017explaining} and layer-wise relevance propagation (LRP) \cite{bharadhwaj2018layer} to learn the attribution score of audio formats in spectrograms. The Gradient-weighted Class Activation Mapping (Grad-CAM) \cite{selvaraju2017grad} is used in \cite{halpern20_odyssey} to identify the significant regions in the spectrogram. Motivated by Grad-CAM, we construct a learnable class activation mechanism to localize the discriminative regions. However, unlike the existing approaches that apply XAI tools externally, our method provides internal justification for the decision-making process by considering both detection capability and outcome interpretability simultaneously.  We utilize the class activation representation within our proposed detection model to identify and visualize the crucial frames that determine detection outcomes.

\section{Proposed Model}
\begin{figure*}[th]
  \centering
  \includegraphics[width=\textwidth]{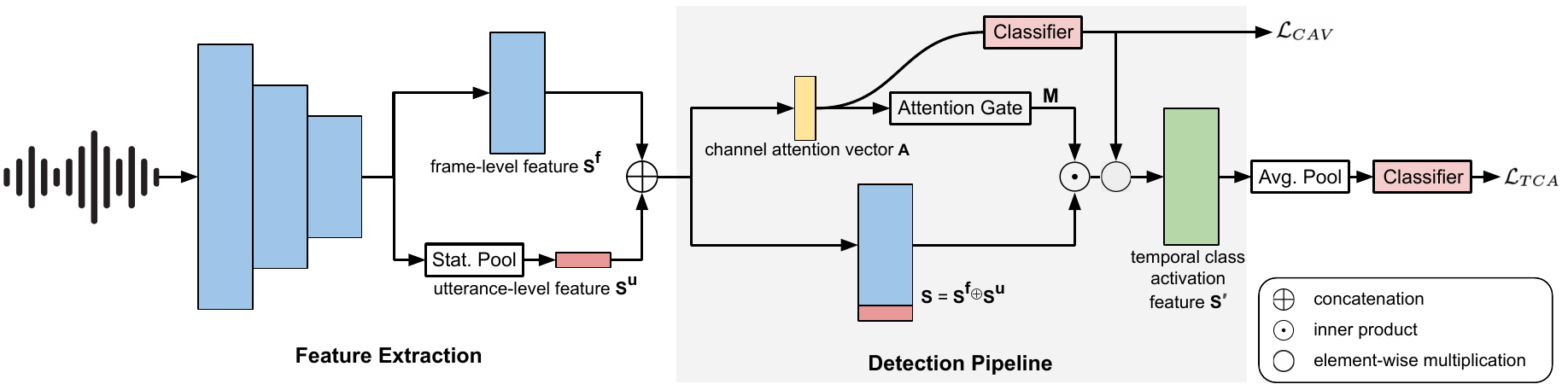}
  \caption{Overall architecture of proposed audio spoofing detection model.}
\label{fig:OVERVIEW}
\end{figure*}

In this section, we elaborate the feature extraction module and the detection pipeline of our  model. The feature extraction module consists of both frame-level and utterance-level representation. The detection pipeline includes channel attention mechanism conditioning on the temporal features. The architecture of the proposed detection model is illustrated in Figure \ref{fig:OVERVIEW}.

\subsection{The SSL-based feature at the frame level}
The SSL models \cite{baevski2020wav2vec,hsu2021hubert} have demonstrated its ability to generate latent representations of raw audio waveform. 
Our proposed model utilizes a pre-trained wav2vec 2.0 XLS-R model \cite{babu2021xls} as the front-end feature extractor to obtain temporal representations for the raw audio inputs. The wav2vec 2.0 model consists of a CNN-based encoder module, a context network with the Transformer architecture, and a quantization module, to produce a quantized latent speech representation that captures the dependent information from the entire audio sequence \cite{baevski2020wav2vec}.  The selected wav2vec 2.0 XLS-R model, with 300 million parameters, is pre-trained on unlabelled speech data from various sources with multiple languages. During the training phase, we fine-tune the all parameters in this pre-trained model with our downstream classifier using labelled training data, which makes this SSL-based front-end feature extractor learn the deep embedding that more adapt to the spoofing detection task. Given an input audio $x$, the corresponding frame-level feature representation $\boldsymbol{S^f} \in \mathbb{R}^{T \times C}$ is extracted, where $T$ and $C$ refer to the number of time frames and channels, respectively. The feature representation is then fed to two stacks consisting of a fully connected (FC) layer, batch normalization (BN) with ReLU activation, and a dropout layer for data downsampling.

\subsection{Attentive statistical feature at the utterance level}
The deep embedding extracted from the wav2vec 2.0 model represent the speech information at frame levels. Additionally, the utterance-level information is also crucial to spoofing detection. Therefore, we implement attentive statistical pooling \cite{wang2018attention} on the frame-level embedding to obtain an utterance-level feature. Given the frame-level embedding, $\boldsymbol{S^f}$,  we first calculate its frame-level attentive score $e_t$ for each frame $t$ by:
\begin{align}
  e_t &= f(W  \boldsymbol{S}^f_t +b),
\end{align}
where $f(\cdot)$ is tanh function, and the parameters $W$ and $b$ are shared across all $C$ channels to avoid overfitting. Then, the score $e_t$ is normalized over all frames using a softmax function:
\begin{align}
  \alpha_t &= \frac{\exp(e_t)}{\sum_{\tau}^T \exp(e_\tau)}.
\end{align}
The normalized attentive score represents the importance of each frame $t$, and then it works as the weights to be applied on the embedding, $\boldsymbol{S^f}$, to calculate the mean and standard deviation respectively, as follows:
\begin{align}
  \tilde \mu &= \sum_{t}^{T} \alpha_t \boldsymbol{S}^f_t,
\end{align}
\begin{align}
  \tilde \sigma &= \sqrt{\sum_{t}^{T} \alpha_t \boldsymbol{S}^f_t \odot \boldsymbol{S}^f_t - \tilde \mu \odot \tilde \mu }.
\end{align}
The weighted mean $\tilde \mu$ and the weighted standard deviation $\tilde \sigma$ are concatenated and projected into 1-D representation  $\boldsymbol{S^u} \in \mathbb{R}^{1 \times C}$ as an utterance-level feature. This weighted statistics describe the distribution of important frames across the entire utterance. In this way, the utterance-level feature provides a higher-level perspective by focusing on specific frames to emphasize discriminative factors in the spoofing process. 

The utterance-level feature $\boldsymbol{S^u}$ and the frame-level feature $\boldsymbol{S^f}$ are concatenated along the time dimension to form the final feature representation $\boldsymbol{S} \in \mathbb{R}^{T' \times C}$, where $T'=T+1$.

\subsection{The detection pipeline using temporal class activation}
Our downstream detection pipeline receives the extracted audio features to determine the type of spoofing while also learning to identify when the spoofing sounds occur, as further detailed below.

\subsubsection{Extracting channel attention vector (CAV)} 
Given the feature embedding $\boldsymbol{S}$, we extract the CAV to indicate the importance of different feature channels contributing each class type $k$ as formulated in Equation \ref{equation:eq1}:
\begin{align}
  \boldsymbol{A}_{k} &= W_k^\top \boldsymbol{S},
  \label{equation:eq1}
\end{align}
where $W_k\in \mathbb{R}^{T'}$ is the weight corresponding to $k$-th class across all time frames, and $\boldsymbol{A}_{k} \in \mathbb{R}^{C}$  is the CAV for each class.

\subsubsection{Classifier on CAV with WCE loss} 
We pass $\boldsymbol{A}_{k}$ through a FC layer to make the first label prediction, resulting in a prediction logit vector $\boldsymbol{z} \in \mathbb{R}^{K}$, where $K$ is the total number of classes. The $\boldsymbol{z}$  and the utterance-level label $\boldsymbol{y}$ are compared to compute a weighted multi-label cross-entropy (WCE) loss in the following formula:
\begin{align}
  \mathcal{L}_{CAV} &= -\frac{1}{K}\sum_{k=1}^{K}\boldsymbol{W}_{CE}[k] \cdot \log \frac{\exp(\boldsymbol{z}[k])}{\sum_{k} \exp(\boldsymbol{z}[k])} \cdot \boldsymbol{y}[k],
  \label{equation:eq2}
\end{align}
where $\boldsymbol{z}[k]$ and $\boldsymbol{y}[k] \in \{0,1\}$ denote the predict logit and ground-truth label of $k$-th class. $\boldsymbol{W}_{CE}$ is the weight assigned to each class $k$.

\subsubsection{Extracting temporal class activation (TCA) feature} We implement a learnable gating mechanism onto the CAV, $\boldsymbol{A}_k$, which effectively selects and emphasizes the discriminative feature channels. The gating mechanism is an FC layer with a softmax function along the dimension of the feature channel, which gives an attentive tensor denoted as $\boldsymbol{M}\in \mathbb{R}^{C \times K}$. Then, we apply $\boldsymbol{M}$ on the feature embedding $\boldsymbol{S}$ through the inner product. Additionally, the prediction logit vector $\boldsymbol{z}$ is used as a class-specific mask, thereby generating an TCA feature $\boldsymbol{S'} \in \mathbb{R}^{T' \times C}$, which highlights the discriminative regions along the temporal domain for each class. We obtain $\boldsymbol{S'}$ as follows,
\begin{align}
\boldsymbol{M}_{c,k} &= \frac{\exp(w_{gate}\boldsymbol{A}_{c,k})}{\sum_{c}^C \exp(w_{gate} \boldsymbol{A}_{c,k})},
\end{align}
\begin{align}
 \boldsymbol{S'} &= \boldsymbol{z} \cdot (\boldsymbol{S} \odot \boldsymbol{M}).
\end{align}
 where $\boldsymbol{A_{c,k}}$ denotes the $c$-th item of the channel attention vector $\boldsymbol{A_k}$, 
 and $w_{gate}$ acts as a scalar weight for the gating mechanism.
\subsubsection{Classifier on TCA feature with WCE loss} The second classifier operates based on the TCA feature $\boldsymbol{S'}$. As $\boldsymbol{S'}$ contains both frame-level and utterance-level information, instead of aggregating all feature elements in $\boldsymbol{S'}$ using global pooling, we separate $\boldsymbol{S'} =\{s'_1, s'_2,...,s'_T, s'_{T+1}\}$ into two segments $\boldsymbol{S'_f} =\{s'_1, s'_2,...,s'_T\}$ of the length of $T$, and $\boldsymbol{S'_u} =\{s'_{T+1}\}$ and then apply average pooling to each feature segment. An FC layer is applied after the pooling operation to obtain a new prediction logit vector $\boldsymbol{z'} \in \mathbb{R}^{K}$ for $\boldsymbol{S'}$. $\boldsymbol{z'}$ is also used to compute a weighted CE loss using Equation \ref{equation:eq2} resulting a $\mathcal{L}_{TCA}$. Both $\mathcal{L}_{TCA}$ and $\mathcal{L}_{CAV}$ utilize the same weight $\boldsymbol{W}_{CE}$ across each class.
 
\subsubsection{Overall objective function} The overall objective function for the detection model is
\begin{align}
  \mathcal{L} &= \lambda_1 \mathcal{L}_{CAV} + \lambda_2 \mathcal{L}_{TCA}
\end{align}
where $\lambda_1$ and $\lambda_2$ are different weight values to balance between two individual losses. $\lambda_1$ and $\lambda_2$ are set to 0.3 and 0.7 respectively in our model.

\section{Experiment and Evaluation}

\subsection{Dataset and evaluation metrics}
We use the ASVspoof2019 logical access (LA) dataset \cite{wang2020asvspoof} for the experiments. The spoofed data in the training and development sets are generated by four TTS methods and two VC methods, while the evaluation set consists of 13 different and unseen methods to evaluate the generalization ability of the detector. We fix all audio samples to the same length of 4 seconds either by truncating the longer audio clips or concatenating the shorter audio clips repeatedly. We evaluate the detection performance with two metrics: minimum normalized tandem detection cost function (min t-DCF) \cite{kinnunen2018t} and the Equal Error Rate (EER). A detection result with a lower min t-DCF value or EER score is regarded to be more accurate. 

       

\begin{table}[th]
  \caption{Performance on the ASVspoof 2019 evaluation set in terms of min t-DCF and pooled EER for state-of-the-art single systems and our proposed system.}
  \label{tab:RESULT}
  \centering
  \begin{tabularx}{\linewidth}{ l| c |c c }
    \toprule
    \textbf{System} & \textbf{Front-end}  & \textbf{min t-DCF}  &\textbf{EER(\%)}      \\
    \midrule
    \textbf{Ours}          &     wav2vec 2.0       &  \textbf{0.0165}  &  \textbf{0.51}          \\
    Ma et al. \cite{ma2022convnext} &     raw waveform    & 0.0187    & 0.64          \\
    Jung et al. \cite{jung2022aasist} &       SincNet      &   0.0275   & 0.83\\
    Li et al. \cite{li4251042multi}  &     SincNet      &     0.0317      &  0.93      \\
    Ma et al. \cite{ma2023boost}    &   LFCC        &    0.0294       &   0.98     \\
    Tak et al. \cite{tak2021end}    &     SincNet      &   0.0335        &  1.06      \\
    Li et al. \cite{li2022role}    &    LFCC       &      0.0345     &   1.06     \\
    Luo et al. \cite{luo2021capsule}&    LFCC       &   0.0328        &   1.07     \\
    Wang et al. \cite{wang2022fully}     &    wav2vec 2.0        &   -        &   1.08     \\
    Wang et al. \cite{wang2021investigating}    &    wav2vec 2.0       &     -     &   1.28     \\
    Yang et al. \cite{yang2023comparative}   &    CQT       &   0.0490        &    1.54    \\
    Hua et al. \cite{hua2021towards}     &   raw waveform       &     -      &   1.64     \\
    Ge et al. \cite{ge21_asvspoof}     &     raw waveform      &    0.0517   &  1.77      \\

    \bottomrule
  \end{tabularx}
  
\end{table}

\begin{table*}[th]
  \caption{Breakdown of EER (\%) performance  for all 13 attacks in ASVspoof2019 LA evaluation set with attack types specified (TTS, VC). Our proposed model, the single state-of-the-art and ablation study results are reported. Pooled EER is shown in the last column.}
  \label{tab:TYPES}
    \centering
    \setlength\tabcolsep{5pt} 
  \begin{tabularx}{\linewidth}{ c | ccccccccccccc | c }
    \toprule
    \multirow{2}{*}{\textbf{System}}& \textbf{A07} & \textbf{A08}& \textbf{A09} & \textbf{A10} & \textbf{A11} & \textbf{A12} & \textbf{A13} & \textbf{A14} & \textbf{A15} & \textbf{A16} & \textbf{A17} & \textbf{A18} & \textbf{A19} & \multirow{2}{*}{\textbf{EER(\%)}}    \\
     &TTS &TTS &TTS &TTS &TTS &TTS &VC &VC &VC &TTS &VC &VC &VC &    \\
    \midrule
    \textbf{Proposed} & \textbf{0.02}& 0.24&0.04&\textbf{0.29}&\textbf{0.06}&0.61&\textbf{0.01}&\textbf{0.10}&1.24&\textbf{0.01}&\textbf{0.61}&0.30&\textbf{0.06} &  \textbf{0.51} \\
     Ma et al. \cite{ma2022convnext} & 0.15& \textbf{0.15}&\textbf{0.02}&0.41&0.10&\textbf{0.06}&0.02&0.35&\textbf{0.41}&0.30&2.19&\textbf{0.27}&0.42 &  0.64 \\
     \midrule
    w/o utterance-level feature  &0.01 &0.02 &0.01 &1.32&0.02  &0.42  & 0.00 & 0.08 & 2.71 & 0.00 & 5.74& 0.61 & 0.04 &1.12 \\
    w/o classifier on CAV   &0.00 & 0.02& 0.00& 0.24& 0.02 &0.06  & 0.01 & 0.06 & 0.83 &   0.00 & 2.01& 1.47 & 0.08 &0.61 \\
    w/ binary label  &0.00& 0.55& 0.01& 0.83&0.04 & 2.57 &0.05  & 0.53 & 9.45 & 0.00 & 14.4 & 1.06 & 0.08 &  4.26 \\

    \bottomrule
  \end{tabularx}  
\end{table*}

\begin{figure*}[th]
  \centering
  \includegraphics[width=\textwidth, height=5.5cm,keepaspectratio]{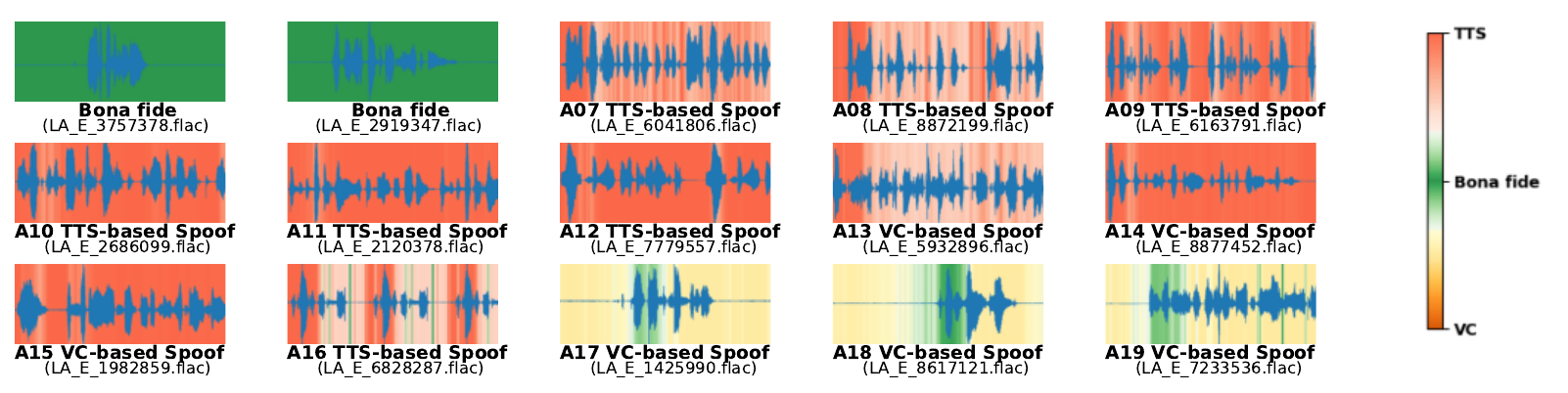}
  \caption{Visualizing the temporal class activation feature on the selected samples, each spoofed audio sample is labeled with its attack type. Different colors denote the class-relevant attention, with color intensity representing the level of contribution to detection results.}
\label{fig:ATTENTION}
\end{figure*} 

\subsection{Model implementation details with multi-label training}
The model is implemented with the PyTorch framework. We adopt the pre-trained XLS-R model \footnote[1]{https://huggingface.co/facebook/wav2vec2-xls-r-300m} with 300 million parameters based on the wav2vec 2.0 base model. The XLS-R model is pre-trained on 436k hours of unlabeled speech in 128 languages, which produces the speech embedding with a dimension size of 1024 in each 20 millisecond. The resulting embedding from the wav2vec 2.0 model is compressed to a size of 128 by two linear layers with 512 and 128 hidden units, with dropout layers set at a rate of 0.2.

During training and validation, we consider spoofing detection as a multi-label classification problem instead of a binary classification. Based on the spoofing generating types, the data labelled as spoofed in the training and validation subsets are categorized into two groups, TTS and VC. Therefore, the ground truth includes three classes of labels in total, which are bonafide, TTS spoofed and VC spoofed. We believe that multi-label training will encourage the model to learn more distinct characteristics to identify TTS and VC-generated speech, thereby potentially increasing the accuracy of detecting spoofing speech.

To manage the data imbalance in the training set, we utilize the WCE loss, where the weights assigned to bonafide, TTS spoofed and VC spoofed are 8, 1, and 1 respectively. An Adam optimizer \cite{kingma2014adam} with a weight decay of $10^{-4}$ is used. The model was trained with 50 epochs with a mini-batch size of 10 and a learning rate of $10^{-5}$. The model with the minimum validation loss for the development set was selected as the best model for evaluation. All experiments were performed on a single GeForce RTX 3090 GPU and the implementation code is publicly available \footnote[2]{https://github.com/menglu-lml/Interpretable-Detection-IS24}.

\subsection{Experiment result}
The performance result of our proposed model is presented in Table \ref{tab:RESULT}. Table \ref{tab:RESULT} also illustrates the performance comparison between our proposed model and the state-of-the-art single systems. The comparisons highlight that our model outperforms not only other single models utilizing the SSL-based features but also End-to-End detection systems and other systems employing a variety of feature types, including learnable feature embedding and hand-crafted acoustic features.

The first two rows of Table \ref{tab:TYPES} presents the breakdown performance of our proposed model and the state-of-the-art for each different spoofing attack in the evaluation set. The results show that our model effectively detects both TTS and VC-based spoofing speech, and outperforms the state-of-the-art method on 8 attacks. In particular, our model achieves a notably low EER score on the A17 attack, which is labelled as the worst-case scenario among all attacks \cite{todisco2019asvspoof}. This is significant because the A17 utilizes direct waveform concatenation method on bonafide human voice, resulting in a greater challenge for detection.

\subsection{Ablation study}
The last three rows of Table \ref{tab:TYPES} illustrates the results of the ablation experiments to demonstrate the merit of the design choices. Removing the utterance-level part in the feature embedding leads to a performance degradation of 54.5\% in terms of EER. 
It shows that the attentive utterance-level feature effectively emphasizes discriminative frames contributing to the detection process across the entire utterance. We demonstrate the underlying connection of CAV to the detection process by ablating the WCE loss upon CAV in the objective function. This leads to a 16\% degradation in EER, dropping to 0.51\%. Notably, with the loss upon CAV, our model enhanceperformances in detecting VC attacks involving the direct conversion of human voices (A17-A19). The effectiveness of multi-label training is also presented. Using binary labels in training results in a degradation to 4.26\% in EER, with the decline primarily attributed to the failure to detect VC attacks. It shows that multi-label training allows the detection model to learn the discriminative factors in TTS and VC-based spoofed audio separately, which gains a deeper understanding of the different characteristics of each attack type.

\subsection{Evaluation of the visual interpretability}
As Figure \ref{fig:ATTENTION} shows, we visualize the temporal class activation feature on the audio samples within the evaluation dataset. The visualization uses different colors to denote the detected audio types for each frame, including TTS-based spoofed, VC-based spoofed, or bonafide. The intensity of color represents the detection confidence. Notably, bonafide and TTS-based spoofed (A07-A12, A16) audio samples are correctly classified in Figure \ref{fig:ATTENTION}. However, some VC-based spoofed samples (A13-A15) are mislabelled as TTS-based, as indicated by the activation feature's color. It occurs because the audio in A13-A15 are generated by the combined VC-TTS spoofing systems, where the TTS voice serves as the source speaker for the VC model. In such cases, our model effectively detects the TTS-based voice source in attacks A13-A15, demonstrating that it has learned the distinct characteristics of spoofed audio generated by TTS and VC. It is further supported by the correct classification of spoofed audio generated by pure VC models that utilize human voice as the source (A17-A19). Additionally, Figure \ref{fig:ATTENTION} localizes the most discriminative frames in the detection process, providing justification for the decision made by our proposed model.

\section{Conclusion}
We are the first to incorporate interpretability directly into the architecture of audio spoofing detection models, enhancing the transparency of their decision-making processes while ensuring a high detection rate. The proposed learnable class activation mechanism identifies and visualizes the crucial frames that contribute detection outcomes. 
Our model achieves an EER of 0.51\% on ASVspoof2019 LA set by leveraging utterance-level features and multi-label classification training. We aim to apply this interpretable model to detect partially spoofed audio by localizing the spoofing segments as future work.


\end{document}